# Referenced Publication Years Spectroscopy applied to *iMetrics: Scientometrics*, *Journal of Informetrics*, and a relevant subset of *JASIST*

Loet Leydesdorff,*[1] Lutz Bornmann,[2] Werner Marx,[3] and Staša Milojević [4]

\* Corresponding author.

[1] loet@leydesdorff.net

Amsterdam School of Communication Research (ASCoR), University of Amsterdam,

Kloveniersburgwal 48, 1012 CX Amsterdam (The Netherlands)

[2] bornmann@gv.mpg.de;

Division for Science and Innovation Studies, Administrative Headquarters of the Max Planck

Society, Hofgartenstr. 8, 80539 Munich (Germany)

[3] w.marx@fkf.mpg.de

Max Planck Institute for Solid State Research, Heisenbergstraße 1, D-70569 Stuttgart (Germany)

[4] smilojev@indiana.edu

School of Informatics and Computing, Indiana University, Bloomington 47405-1901 (United

States of America).

**Abstract**

We have developed a (freeware) routine for "referenced publication years spectroscopy" (RPYS) and apply this method to the historiography of "*iMetrics*," that is, the junction of the journals *Scientometrics, Informetrics*, and the relevant subset of *JASIST* (approx. 20%) that shapes the intellectual space for the development of information metrics (bibliometrics, scientometrics, informetrics, and webometrics). The application to information metrics (our own field of research) provides us with the opportunity to validate this methodology, and to add a reflection about using citations for the historical reconstruction. The results show that the field is rooted in individual contributions of the 1920s-1950s (e.g., Alfred J. Lotka), and was then shaped intellectually in the early 1960s by a confluence of the history of science (Derek de Solla Price), documentation (e.g., Michael M. Kessler's "bibliographic coupling"), and "citation indexing" (Eugene Garfield). Institutional development at the interfaces between science studies and information science has been reinforced by the new journal *Informetrics* since 2007. In a concluding reflection, we return to the question of how the historiography of science using algorithmic means—in terms of citation practices—can be different from an intellectual history of the field based, for example, on reading source materials.

**Highlights**

- Referenced Publication Years Spectroscopy (RPYS) is used to show the historical origins of *iMetrics* (information metrics; bibliometrics; scientometrics) in scholarly literature;

- Whereas Lotka (1926) is indicated as the first source, the intellectual program of *iMetrics* was shaped only in the early 1960s;

- The *Journal of Informetrics* has bridged research traditions in quantitative science studies (*Scientometrics*) and the information sciences (*JASIST*);

- RPYS is routinized and a program is made available at the Internet (at http://www.leydesdorff.net/software/rpys).

# 1    Introduction

In his Postscript to the second edition of *The Structure of Scientific Revolutions*, Kuhn ([1970] 2012) argued that "the key to understanding the structure of research communities was to draw on the recent work in the sociology of science" (Wray, 2013: 78). Among others, Kuhn (2012: 177) refers to the studies of Derek de Solla Price (Price, 1965) and Eugene Garfield (Garfield, 1964), who used bibliometric data (from the newly created *Science Citation Index*) for the historiography of scientific developments. According to Wray (2013), the use of bibliometric data can even be considered as "the cutting edge sociology of science" that may "have the potential to enrich our understanding of the nature and structure of scientific research communities" (p. 78). This potential has already been studied by bibliometricians (e.g., Garfield & Sher, 1963; Garfield *et al.*, 2003; Leydesdorff, 2010). Marx and Bornmann (2010 and 2013), for example, studied the transition from a static to a dynamic conceptualization of the universe in cosmology, and the emergence of plate tectonics in geology using bibliometric data (Bornmann & Marx, 2012).

Developed by Eugene Garfield—who was assisted later by Alexander Pudovkin—the computer program HistCite™ "facilitates the understanding of paradigms by enabling the analyst to identify the significant works on a given topic" (Garfield, Pudovkin, & Istomin, 2003:400).[1] Lucio-Arias and Leydesdorff (2008) enriched the output of HistCite™ with algorithms from social network analysis and information theory. Recently, Marx *et al.* (in press) have introduced a quantitative method named Referenced Publication Years Spectroscopy (RPYS).[2] With this method one can map the historical roots of research fields and quantify their impact on current research. RPYS is based on analyzing the frequencies with which references are cited in the publications of a specific research field (or any other publication set) in terms of the publication years of the cited references.

---

[1] HistCite can be downloaded for free at http://interest.science.thomsonreuters.com/forms/HistCite/ .

The results of Marx *et al*. (in press) were based on the installation of the Science Citation Index (SCI) in the SCISEARCH database at STN International (http://www.stn-international.com/). This installation allows the user to download specifically the frequencies with which references are cited in the publications aggregated to referenced publication years (RPYs). However, access to this version of SCI is uncommon among bibliometricians. In this study, we introduce a computer program which allows for the routinization of RPYS on the basis of data downloaded from the Web of Science (WoS, Thomson Reuters). The program is tested in this study for showing the historical roots of the specialty of *iMetrics* as an example.

*iMetrics* or "information metrics" was introduced as a label by Milojević and Leydesdorff (2013: 141). These authors argue that ""(b)ibliometrics', 'scientometrics', 'informetrics', and 'webometrics' can all be considered as manifestations of a single research area with similar objectives and methods, which we call 'information metrics' or *iMetrics*." Three journals were identified as core journals for *iMetrics* research: *Scientometrics*, a subset of the articles in *Journal of the American Society for Information Science and Technology* (*JASIST*), and the *Journal of Informetrics* (*JoI*). We follow the procedure of Milojević and Leydesdorff (2013) for distinguishing the *iMetrics* literature as an emerging specialty and apply RPYS to both these three (sub)sets and their aggregate.[3]

## 2       Referenced Publication Years Spectroscopy (RPYS)

Conventional citation analysis is based on document sets comprising for example the publications of a researcher, a research group, a research institution, or a journal. The number of times these publications are *cited* across the database is analyzed in order to evaluate research performance. In the context of mapping aggregated citation relations for consecutive years—for example, among journals—Leydesdorff and Cozzens (1993) argued in favor of focusing instead on *citing* behavior for modeling the dynamics of science, because "citing" represents the current variable whereas the "cited" dimension refers to the archive of science. Similarly, one can distinguish at the document level between co-citation analysis (Marshakova, 1973; Small, 1973) and bibliographic coupling by the citing documents (Kessler, 1963). The two dimensions (citing and cited) provide us with different perspectives on the structure and dynamics of science (Leydesdorff, 1993).

Bornmann and Marx (2013) proposed to use the perspective of citing in order to show that one can thus limit citation analyses to specific research fields by first selecting the relevant publications and then analyzing the references cited within this domain. Thus, one can determine the impact of publications, authors, institutions, or journals within a specific research field from the perspective of hindsight. Cited references can be used to study an author's intellectual history

(White 2001) and can constitute "a form of watermark for their scholarly output" (Cronin & Shaw, 2002). Collectively, references to prior literature can also be viewed as "a vital piece in the collaborative construction of new knowledge between writers and readers" (Hyland, 2004, p. 21).

The analysis of cited references with specific emphasis on the publication years of these references can be used to quantify the significance of historical publications, and to reveal the historical roots of a given research field. The distribution of the cited publications over their publication years (the "referenced publication years," RPYs, not to be confused with the method RPYS) is typically at a maximum a few years before the publication year of the citing publications and then tails off significantly into the past. For example the distribution of reference ages for the three *iMetrics* journals was shown to peak at 2-4 years (Milojević & Leydesdorff 2013). However, this differs among fields of science (Price, 1970).

In fields with fast moving research fronts, most references refer to recent specialist literature in the same domain of literature in which the citing publications have appeared; only a relatively small proportion of the cited publications is older or points to other disciplines.[4] The (often steep) decline over time is associated with the fact that specialist literature at the research front becomes less interesting and important as time passes because of ageing. The relative decline can also be the result of an abrupt acceleration of the increase in specialist literatures in all disciplines which began around 1960 ("Sputnik shock") and continues to this day (Marx, 2011; cf. Althouse *et al.*, 2009). Different disciplines exhibit different average ages of references, ranging from 8 to 18 years; this average age is driven almost entirely by variations in the fraction of references in the paper made to foundational work that is more than ten years old (Milojević, 2012).

---

[4] Using the routines crciting.exe and crcited.exe (available at http://www.leydesdorff.net/journals11), one can also compare Rao-Stirling diversity measures for interdisciplinarity across sets of citing or cited references (Leydesdorff *et al.*, 2013).

More detailed analysis of the publication years of all the references cited in specific research fields has shown that RPYs lying further back in the past (at the beginning of the 20[th] century and earlier) are not represented equally, but that some RPYs appear pronouncedly in the references. These frequently occurring RPYs become more differentiated towards the past and mostly show distinct peaks in the RPY distribution curves. If one analyzes the publications underlying these peaks in the 19[th] and the first half of the 20[th] century, these often contain only single highly-cited publications in a specific year. Towards the present, the peaks of individual publications lie over a broad continuum of newer publications and are less pronounced. Due to the large number of publications cited at the research front, the proportion of individual highly-cited publications within a specific RPY can be expected to decline rapidly in the more recent past.

The focus on the most frequently cited publications in the history of a specific research field provides a special application of cited-reference analysis. In analogy to the spectra in the natural sciences, which are characterized by pronounced peaks in the quantification of certain properties (such as the absorption or reflection of light as a function of chemical structures), this application was called "referenced publication years spectroscopy" (RPYS) by Marx *et al*. (in press). RPYS reveals the historical papers (potentially) most relevant for the evolution of a specific research field which could (or should?) be taken into consideration when discussing its history. However, their historical role can only be determined in a careful analysis by experts in the field under study. We turned to our own emerging field that was characterized recently by two of us as *iMetrics* (Milojević & Leydesdorff, 2013), for the purpose of this validation.

## 3 Methods

### 3.1 Data

For the investigation of the historical roots of *iMetrics*, we downloaded from the WoS 9,244 papers (all document types) published in the following three journals: *Journal of the American Society for Information Science and Technology* (*JASIST*), *Scientometrics,* and *Journal of Informetrics* (*JoI*) (date of search: July 21, 2013). The search was not restricted to a specific time period. These three journals contain the core of *iMetrics* papers (Bradford, 1934). Milojević & Leydesdorff (2013) found that *Research Evaluation* (the journal with the largest number of articles on the topic after the three journals we included) has two times fewer *iMetrics* articles that JoI, the smallest of the core journals.

Among these three journals, *JASIST* has the longest tradition: before 1970 it was issued under the titles of *Journal of Documentary Reproduction* (until 1942) and *American Documentation* (until 1970). Until 2001, the journal was published as *Journal of the American Society for Information Science*, but the title was changed to the current one in 2002.[5] *Scientometrics* started in 1978. *JoI* is the youngest among the three *iMetrics* journals; it has been published since 2007. In accordance with these time periods of being active, the numbers of papers and cited references in

---

[5] The name of this journal will shortly be changed to *Journal of the Association for Information Science and Technology*.

Table 1 show that at the time of our downloads *JASIST* had the highest and *JoI* the lowest numbers in terms of both papers and cited references.

**Table 1**: Number of papers and number of cited references in three journals relevant for *iMetrics*.

| Journal | Number of papers | Number of cited references |
|---|---|---|
| *Journal of the American Society for Information Science and Technology* (since 1970) | 5,319 | 123,872 |
| *iMetrics* **subset of *JASIST* (*JASIST-I* since 1982)** | **782** | **27,530** |
| Other subset of *JASIST* since 1982 (*JASIST-O*) | 3,609 | 87,565 |
| *Scientometrics* **(since 1978)** | **3,547** | **79,593** |
| *Journal of Informetrics* **(since 2007)** | **378** | **11,887** |
| **Total *iMetrics*** | **4,707** | **119,010** |
| Total | 9,244 | 213,352 |

Because the *JASIST* set is a mixture of papers in *iMetrics* and other domains in the library and information sciences (Lucio-Arias & Leydesdorff, 2009), Milojević and Leydesdorff (2013) tested a routine to distinguish between the two sets using two criteria: specific title words and at least one reference to either *Scientometrics* or *JoI*. However, the first references to *Scientometrics* appeared in *JASIST* only in 1982, and therefore we could decompose the sets in terms of papers only from this date onwards. Thus, of the 5,319 *JASIST* documents, we classify with high precision only the 4,391 articles published since 1982 using this routine: 782 *JASIST* papers being classified as *iMetrics* and 3,609 as non-*iMetrics* papers.

The *iMetrics* subset of *JASIST* will be denoted as *JASIST*-I. The papers in *JASIST*-I contain 27,530 cited references. (Table 1 provides the descriptive statistics.) The non-*iMetrics* papers will be denoted as *JASIST*-O; they contain 87,565 cited references. The percentage of *iMetrics* papers in *JASIST* has been increasing steadily since 1995 with an accelerated rate of growth since 2005.

Figure 1 shows the development of the *JASIST*-I set also as a percentage of the larger set of *JASIST*.

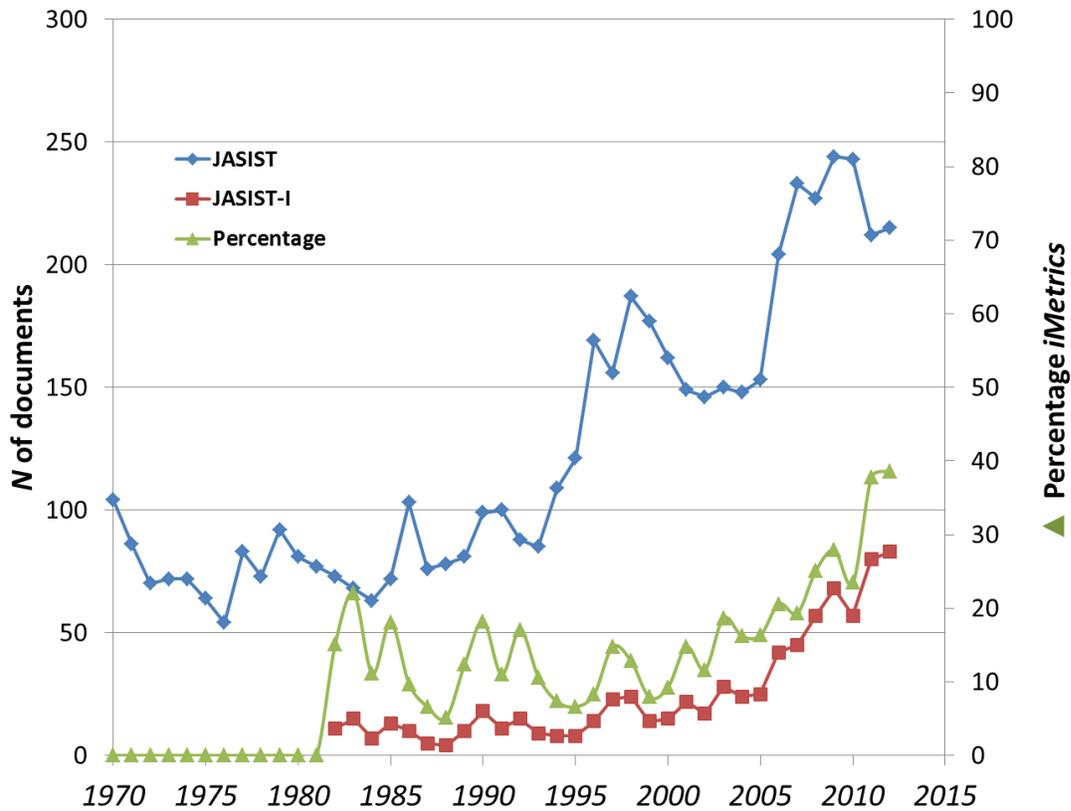

**Figure 1**: Subset of *iMetrics* papers (■) in *JASIST* (♦); as a percentage (▲).

### 3.2   The program RPYS.exe

The program RPYS.exe can be used to generate a RPYS of any set downloaded from the WoS. The procedure of how to use the routine is described in detail at

http://www.leydesdorff.net/software/rpys/. (The freeware routine can also be downloaded from this page.) Based on WoS data as input (for example, papers published in the three *iMetrics* journals in this study), the program generates two output files: "rpys.dbf" and "median.dbf."

"Rpys.dbf" organizes the number of cited references per referenced publication year. This file can be used in Excel for drawing a spectrogram of the data. "Median.dbf" contains the deviation of the number of cited references in each year from the median for the number of cited references in the two previous, the current, and the two following years ($t - 2; t - 1; t; t + 1; t + 2$). This deviation from the five-year median provides a curve smoother than the one in terms of absolute numbers. Both curves can be visualized using median.dbf (e.g., in Excel). Figures 2 – 7 below will provide examples.

## 4     Results

### 4.1.    *iMetrics*: the aggregated set

The distribution of the number of references cited in the *iMetrics* literature (*JASIST*-I, *Scientometrics*, and *JoI* combined) across the publication years is shown in Figure 2. As noted, the citing *iMetrics* literature has been published between 1982 and the present, whereas the time window of the cited publications—the references cited within the citing *iMetrics* publications analyzed here—extends from 1900 to 1970 in order to focus on historical publications. We did not include pre-1900 references because these were far less numerous, more erroneous, and also less important in our relatively young field (compared to physics, for example). Post-1970 references were not included because we are interested in the historical roots of *iMetrics*. This cut-out makes the distinct peaks of the most frequently cited historical publications clearly visible (Marx *et al.*, in press).

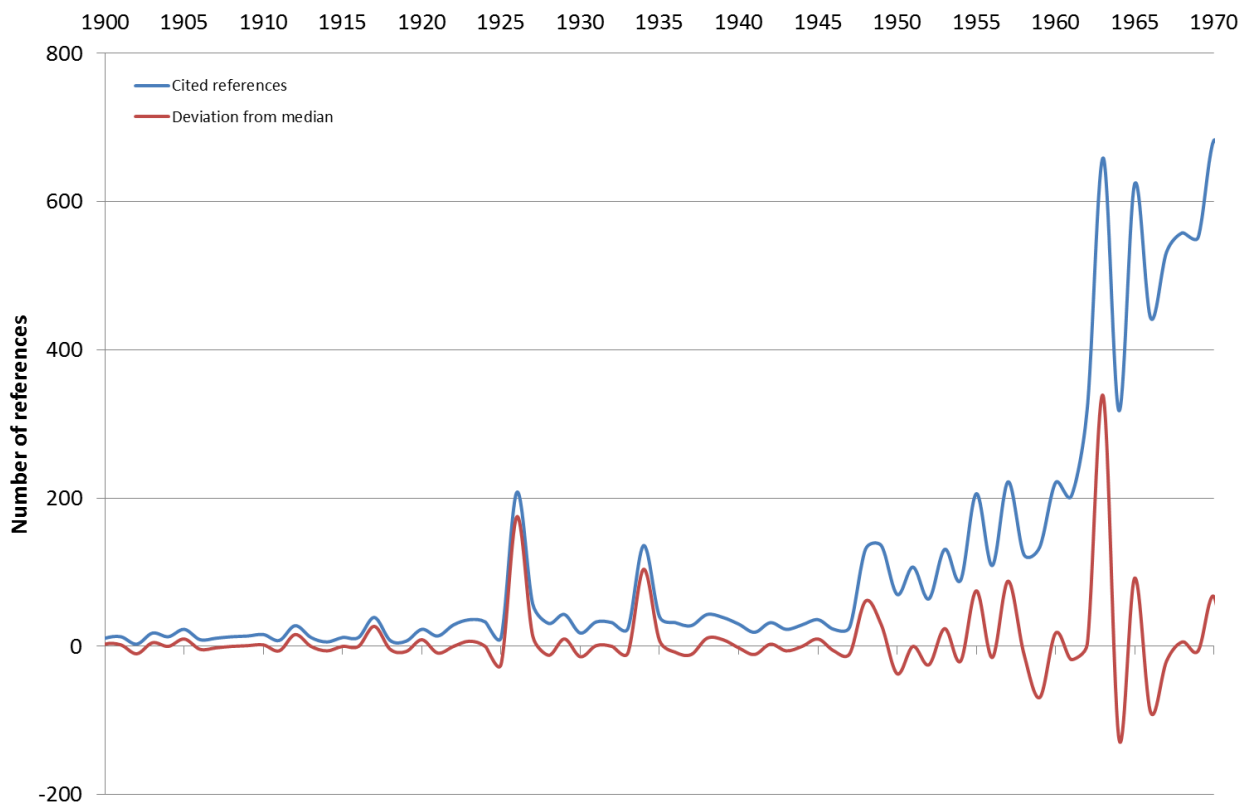

Publication years with the three most significant peaks:

1. 1963: of the 659 cited references in total:
   - 213 refer to Price, D. J. de Solla. (1963). *Little science, big science*. New York: Columbia University Press.
   - 82 refer to Kessler, M.M. (1963). Bibliographic coupling between scientific papers. *American Documentation, 14*(1), 10-25.
   - 41 refer to Garfield, E., & Sher, I. (1963). New factors in the evaluation of scientific literature through citation indexing. *American Documentation*, *14*(3), 195-201.
   - 17 refer to Garfield, E. (1963). Citation indexes in sociological and historical research. *American Documentation*, *14*(4), 289-291.

2. 1926: of the 208 cited references in total:
   - 184 refer to Lotka, A.J. (1926). The frequency distribution of scientific productivity. *Journal of the Washington Academy of Science*, *16*(12), 317-323; in seven additional documents this reference is incomplete.

3. 1934: of the 136 cited references in total:
   - 99 refer to Bradford, S. C., (1934). Sources of information on specific subjects. *Engineering*, *137*, 85-86; in seven additional documents this reference is incomplete.

**Figure 2**. Referenced Publication Years Spectroscopy (RPYS) of three *iMetrics* journals: *Scientometrics*, *JASIST*-I, and *Journal of Informetrics*; *N* of documents = 4,708; *N* of citations = 119,010.

In addition to the distribution of the cited references, Figure 2 shows also the deviation of the number of cited references in each year from the median of the numbers of cited references in the two previous, the current, and the two following years (the deviation from the 5-year median). It is particularly easy to see the peaks created by the frequently cited historical publications using this (second) curve, since the deviation from the median corrects for the upward trend in more recent years (among other things).

The largest peaks in Figure 2 are visible for the years 1963 (with 659 cited references), 1926 (with 208 cited references), and 1934 (with 136 cited references). The historiography indicated by these peaks will be discussed in section 4.4, that is, after the decomposition.

## 4.2    *iMetrics* decomposed

Let us first analyze the RPYS of the three individual journals: *Scientometrics*, *JASIST-I,* and *JoI*. Since our model is that of a junction among the three journals shaped increasingly during the 1990s and 2000s (Lucio-Arias & Leydesdorff, 2009), it is interesting to study whether different historical roots were imported from the various origins. Note that Van den Besselaar and Leydesdorff (1996) analyzed the emergence of artificial intelligence around 1988 as a grouping between three journals (*Artificial Intelligence*, *AI Magazine,* and *IEEE Expert*)[6] that were constitutive for this field. It can be shown that the configuration of three different components provides a structural condition for the auto-catalysis and potential emergence of a new

---

[6] This journal was renamed as *IEEE Intelligent Systems* in 2001 (after first having been renamed *IEEE Intelligent Systems & Their Applications* in 1997).

component (Ivanova & Leydesdorff, in preparation; Leydesdorff & Ivanova, in press; Ulanowicz 2009).

In the case of *JASIST*-I, this "journal" is a subset within the larger set of *JASIST*; but the structural communality between *Scientometrics* as a European journal with a constituency different from *JASIST* was perhaps brought to the fore by the emergence of *JoI* in 2007. *JoI* programmatically linked *Scientometrics* with the information sciences, and followed upon the establishment of the International Society for Scientometrics and Informetrics (ISSI) in 1991. Furthermore, Eugene Garfield—who was among the founding fathers of *Scientometrics*—was elected President of the ASIST in 1999. Analyzing the similarities of the knowledge bases of these three venues, Milojević & Leydesdorff (2013) concluded that *JoI* can be considered as bridging between *JASIST-I* and *Scientometrics*.

*Scientometrics*

Not surprisingly—because it is the largest subset of *iMetrics* (66.9%)—the RPYS for *Scientometrics* (Figure 3) is in many respects very similar to the RPYS of *iMetrics* as a whole (Figure 2). The peaks in 1926, 1934, and 1963 are virtually the same. However, the RPYS for *Scientometrics* shows an additional peak in 1957. Interestingly, the latter peak is more important than the one of 1934, both in terms of absolute numbers and deviance from median values. Furthermore, the authors cited at this peak are characteristic for *Scientometrics* and not for the other two venues: Merton as an author in the sociology of science and Farrell as a quantitative economist focusing on technology and innovation.

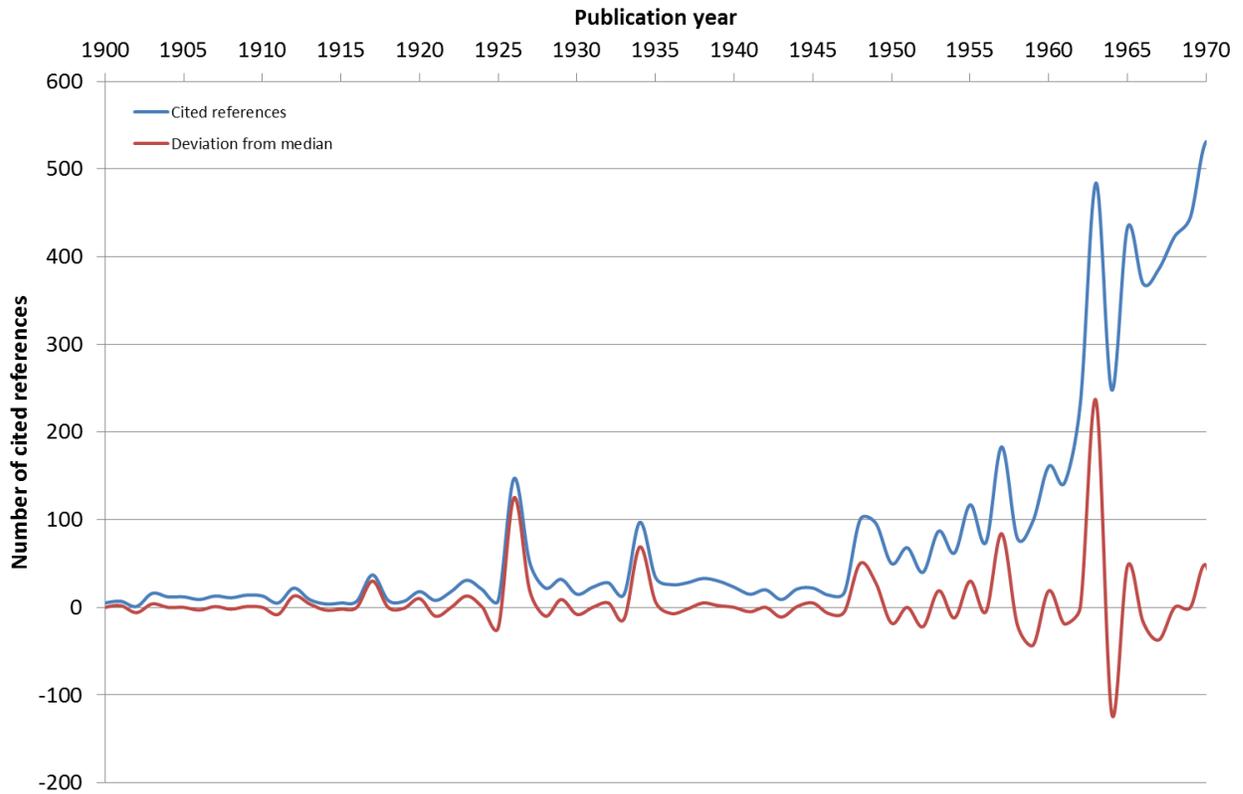

Publication years with the three most significant peaks:

1. 1963: of the 484 cited references in total:
   - 212 refer to Price, D. J. de Solla. (1963). *Little science, big science*. New York: Columbia University Press.
   - 50 refer to Kessler, M.M. (1963). Bibliographic coupling between scientific papers. *American Documentation, 14*(1), 10-25.
   - 38 refer to Garfield, E. (1963). Citation indexes in sociological and historical research. *American Documentation*, *14*(4), 289-291.
2. 1926: of the 147 cited references in total:
   - 132 refer to Lotka, A.J. (1926). The frequency distribution of scientific productivity. *Journal of the Washington Academy of Science*, *16*(12), 317-323.
3. 1957: of the 167 cited references in total:
   - 19 refer to Merton, R.K. (1957). Priorities in scientific discovery: A chapter in sociology of science. *American Sociological Review*, *22*(6), 635-639.
   - 12 refer to Farrell, M.J. (1957). The measurement of productive efficiency. *Journal of the Royal Statistical Society. Series A (General), 120*(3), 253-290.

**Figure 3**. Referenced Publication Years Spectroscopy (RPYS) of *Scientometrics* (*N* = 3,547 documents with 79,593 cited references).

*JoI* is the smallest and youngest contributor to the *iMetrics* field, and, as noted, its function can be seen as mediating between the intellectual programs in the other two contexts of science studies and the information sciences. Figure 4 shows the RPYS for this journal. It shares with the previous RPYS, the peaks in 1926 and 1934. However, its most prominent peaks are in 1955, 1963, and 1968. The peak of 1963 coincides with those of *Scientometrics* (Figure 3) and *JASIST-I* (Figure 5). However, the peaks in 1968 and 1955 are not as prominent in the other two venues.

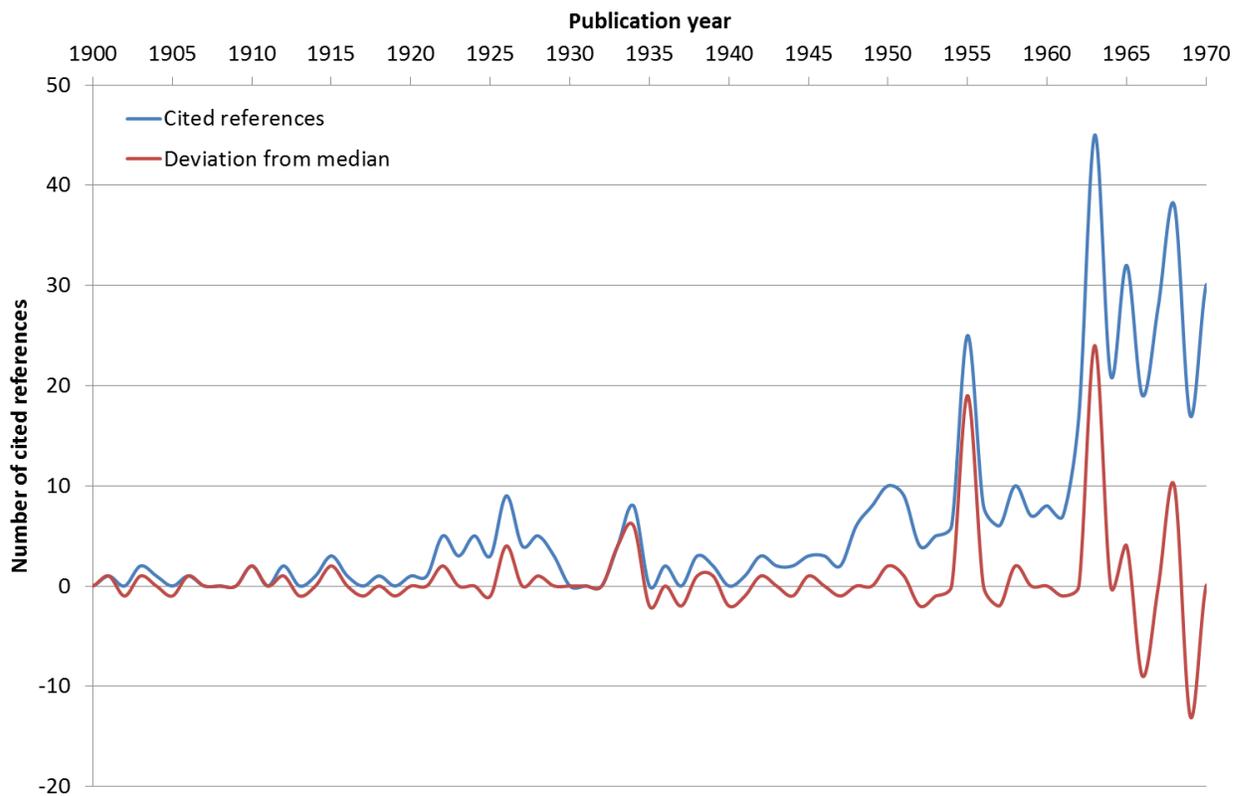

Publication years with the three most significant peaks:
1. 1963: of the 45 cited references in total:
   - 15 refer to Price, D. J. de Solla. (1963). *Little science, big science*. New York: Columbia University Press;
   - 13 refer to two papers of Kessler:
   - Kessler, M.M. (1963). Bibliographic coupling between scientific papers. *American Documentation, 14*(1), 10-25;
   - Kessler, M.M. (1963). Bibliographic coupling extended in time: Ten case histories. *Information Storage & Retrieval, 1*(4), 169-187.
2. 1955: of the 25 cited references in total:

- 18 refer to Garfield, E. (1955). Citation indexes for science: A new dimension in documentation through association of ideas. *Science, 122*(3159), 108-111.
3. 1968: of the 38 cited references in total:
   - 20 refer to Merton, R.K. (1968). The Matthew effect in science: The reward and communication systems of science are considered. *Science, 159*(3810), 56-63.

**Figure 4**: Referenced Publication Years Spectroscopy (RPYS) of the *Journal of Informetrics* (*N* = 378 documents with 11,887 cited references).

*JASIST-I*

*JASIST-I* is a subset of *JASIST*; however, it is more than twice the size of *JoI*, in terms of both numbers of publications and cited references. Figure 5 provides the RPYS of this data set. It shares with the previous RPYS the peaks in 1926 and 1963. Its most prominent peaks are in 1963 and 1965. The peak of 1963 coincides with both *Scientometrics* (Figure 3) and *JoI* (Figure 4). Its 1965 peak is not so prominent as in the other two venues.

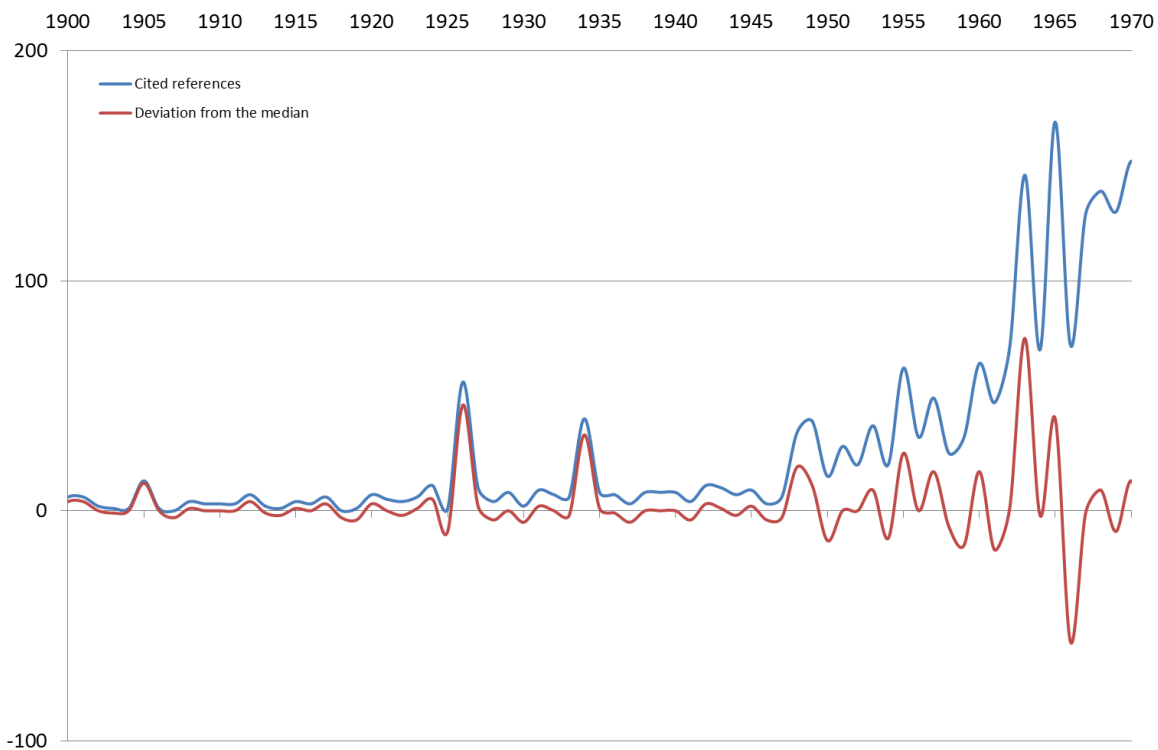

Publication years with the three most significant peaks:

1. 1963: of the 146 cited references in total,:
   - 40 refer to Price, D. J. de Solla (1963). *Little science, big science*. New York: Columbia University Press;
   - 30 refer to two papers by Kessler:
     Kessler, M.M. (1963a). Bibliographic coupling between scientific papers. *American Documentation, 14*(1), 10-25 ;
     Kessler, M.M. (1963b). Bibliographic coupling extended in time: Ten case histories. *Information Storage & Retrieval, 1*(4), 169-187;
   - 13 refer to Garfield, E., & Sher, I. (1963). New factors in the evaluation of scientific literature through citation indexing. *American Documentation, 14*(3), 195-201.

2. 1965: of the 169 cited references in total:
   - 50 refer to Price, D. J. de Solla. (1965). Networks of scientific papers. *Science, 149*(3683), 510-515;
   - 16 refer to Kaplan, N. (1965). The norms of citation behavior: Prolegomena to the footnote. *American Documentation, 16*(3), 179-184.

RPY 1926: of the 56 cited references in total:
   - 55 refer to Lotka, A.J. (1926). The frequency distribution of scientific productivity. *Journal of the Washington Academy of Science, 16*(12), 317-323.

**Figure 5**. Referenced Publication Years Spectroscopy (RPYS) of the *iMetrics* subset of the *Journal of the American Society for Information Science and Technology* (*JASIST*-I; *N* = 782 documents with 22,530 cited references.

## 4.3    *JASIST and JASIST-O* as relevant contexts

*JASIST*

*JASIST* is one of the major journals in the field of information science. A number of studies considered *iMetrics* as having become an integral part of information science (e.g., van den Besselaar 2001). However, Milojević & Leydesdorff (2013) conclude that *iMetrics* researchers are using *JASIST* more as an additional venue for sharing *iMetrics* research than for participating in the field of information science. Thus, it is informative to examine the historical roots first of *JASIST* as a whole, and then also the *JASIST* subset that publishes information science other than *iMetrics* papers: *JASIST*-O.

Figure 6 shows the spectrogram for *JASIST* before the decomposition. This set (*N* = 5,319) includes both the 782 papers that were selected as *iMetrics* as a subset and the other papers. As noted, the *iMetrics* papers contain on average more cited references than the other papers in *JASIST*, yet they remain a minority in total (14.7% of the publications containing 22.2% of the cited references; see Figure 1).

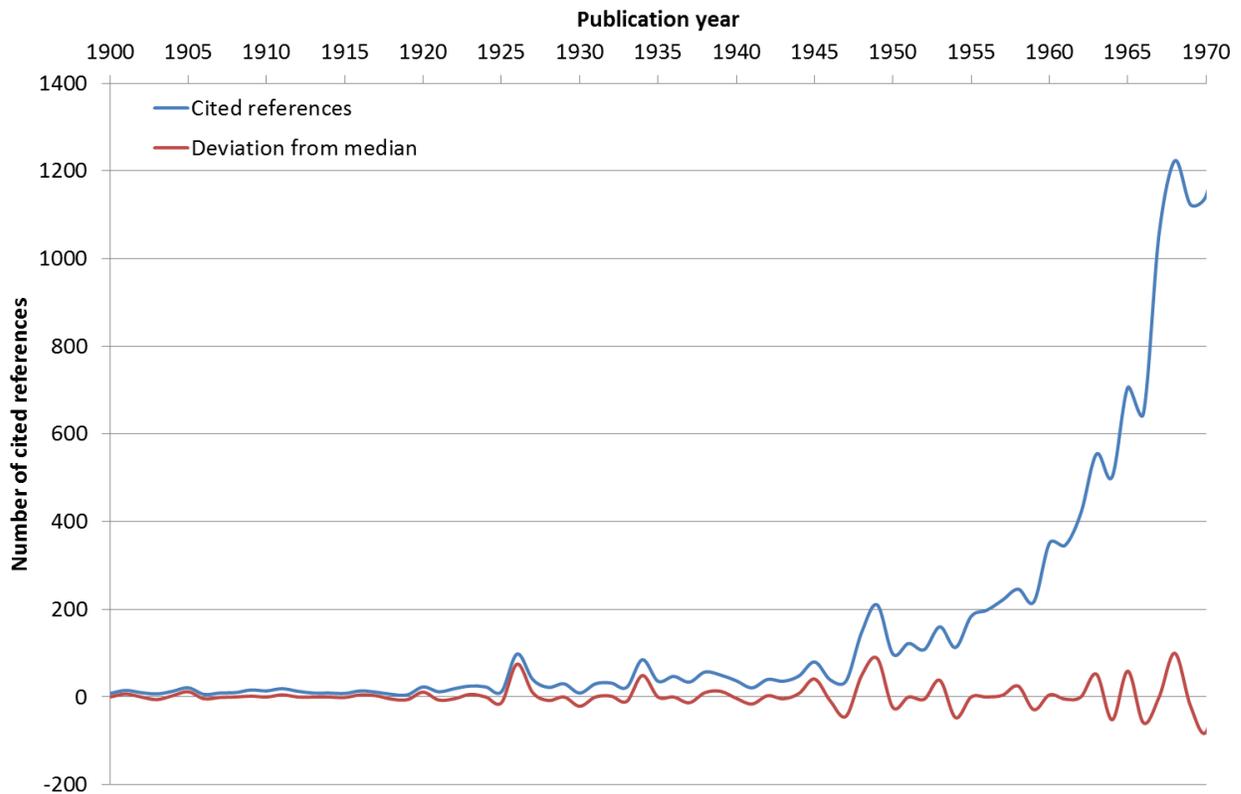

Publication years with the three most significant peaks:
1. 1968: of the 1,223 cited references in total:
   - 63 refer to Salton, G. (1968). *Automatic information organization and retrieval.* New York: McGraw-Hill;
   - 59 refer to Taylor, R.S. (1968). Question negotiation and information seeking in libraries. *Journal of College and Research Libraries, 29*(3), 178-194;
   - 45 refer to Merton, R.K. (1968). The Matthew effect in science: The reward and communication systems of science are considered. *Science, 159*(3810), 56-63.
2. 1949: of the 210 cited references in total:
   - 65 refer to Zipf, G.K. (1949). *Human behavior and the principle of least effort.* Reading, MA: Addison-Wesley;
   - 63 refer to Shannon, C.E., & Weaver, W. (1949). *The mathematical theory of communication.* Urbana: University of Illinois Press.
3. 1926: of the 98 cited references in total:
   - 84 refer to Lotka, A.J. (1926). The frequency distribution of scientific productivity. *Journal of the Washington Academy of Science, 16*(12), 317-323.

**Figure 6**. Referenced Publication Years Spectroscopy (RPYS) of the *Journal of the American Society for Information Science and Technology*; all documents (*N* = 5,319) with 123,872 cited references.

The two peaks of 1926 and 1934 are also visible in this spectrogram. This is not surprising, given the relatively large number of *iMetrics* papers published in *JASIST*. Focusing on the three main peaks, however, we see that 1934 is no longer one of them. Rather, 1968 and 1949 are more important in absolute terms. The papers in the 1949 and 1968 peaks are not so prominent in any of the *iMetrics* datasets we have examined above. These peaks rather highlight citation classics of the other fields relevant to the library and information sciences, including works by Gerald Salton, Claude E. Shannon, George K. Zipf, and Robert S. Taylor. The 1963 peak that we identified as pivotal for the development of *iMetrics* by being present in all three venues and thus probably constitutive for the specialism is not shared in this environment.

*JASIST-O*

For a more detailed view of the specific historical roots of information science in comparison to those of *iMetrics*, we analyzed also the *JASIST*-O subset of *JASIST*. Figure 7 provides the RPYS of this data set. The largest peaks are now in the years 1968 (with 530 cited references), 1967 (with 463 cited references), and 1949 (with 132 cited references). All of these peaks are unique to this dataset and do not appear in any of the *iMetrics* ones. Furthermore, none of the works cited in these peaks appear in any of the *iMetrics* sets.

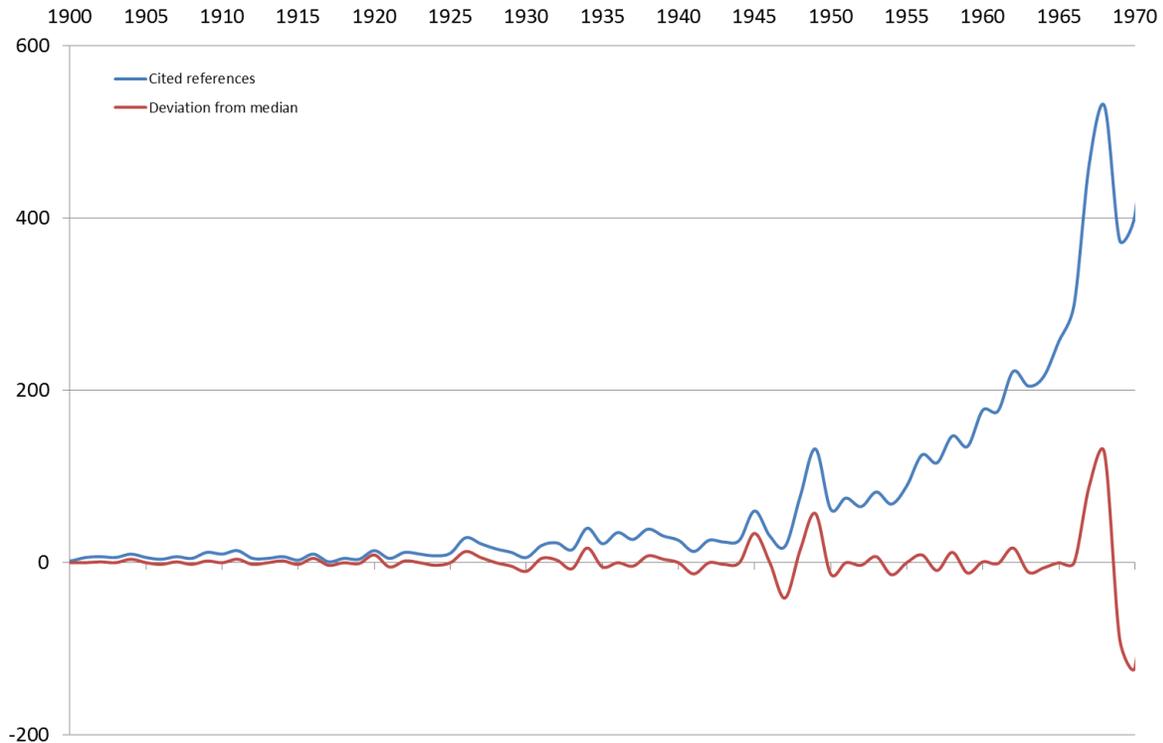

Publication years with the three most significant peaks:
1. 1968: of the 530 cited references in total:
   - 52 refer to Taylor, R.S. (1968). Question negotiation and information seeking in libraries. *Journal of College and Research Libraries, 29*(3), 178-194
   - 27 refer to Salton, G. (1968). *Automatic information organization and retrieval.* New York: McGraw-Hill.
   - 25 refer to Lovins, J.B. (1968). Development of a stemming algorithm. *Mechanical Translation and Computational Linguistics*, *11*, 22-31.
2. 1967: of the 463 cited references in total:
   - 32 refer to Glaser, B., & Strauss, A. (1967). *The discovery of grounded theory: Strategies for qualitative research.* Chicago: Aldine Publishing Co.;
   - 22 refer to Cuadra, C.A., & Katter, R. (1967). Opening the blackbox of 'relevance'. *Journal of Documentation*, *23*(4), 291-303;
   - 14 refer to Rees, A.M. & Schultz, D.G. (1967). *A field experiment approach to the study of relevance assesments in relation to document searching: Final report. Volume 1.* Cleveland, Oh: CWRU.
3. 1949: of the 132 cited references in total:
   - 42 refer to Shannon, C.E., & Weaver, W. (1949). *The mathematical theory of communication.* Urbana: University of Illinois Press;
   - 26 refer to Zipf, G.K. (1949). *Human behavior and the principle of least effort.* Reading, MA: Addison-Wesley.

**Figure 7**. Referenced Publication Years Spectroscopy (RPYS) of the non-*iMetrics* subset of the *Journal of the American Society for Information Science and Technology* (*JASIST*-O; *N* = 3,609) documents with 87,565 cited references.

The major works cited here attest to the importance for the field in general of the mathematical theory of communication by Shannon and Weaver (1949).[7] In addition to a number of works of importance to information retrieval (e.g., an early book by Salton [1968] and Lovins [1968]), Cuadra and Katter (1967) and Rees and Schultz (1967) focus on "relevance" as a crucial concern in information retrieval. The importance of information-seeking behavior as another major research topic within information science is illustrated by the presence of early models both within the field (e.g., Taylor [1968]) and from outside this domain (e.g., Zipf [1949]). The references to Glaser and Strauss (1967) show the increased importance of qualitative methods in information-science research, since this title has become the classical reference to "grounded theory" (that is, in-depth case studies) in the social sciences.

These findings pointing to the differences in historical roots between *iMetrics* and information science are an additional indicator that these are two separate research fields.

## 4.4. *iMetrics*: Summary and Conclusions

Our analyses have identified a number of works that form the foundation of the field of *iMetrics*. These works were published between 1926 and 1968. The references to publications in 1926 are almost exclusively (92%) to a single publication, namely to Alfred J. Lotka's study entitled "The frequency distribution of scientific productivity" in the *Journal of the Washington Academy of Sciences* (Lotka, 1926). This is one of the few publications that were among the top influencers in all three venues.

---

[7] The mathematical theory of communication was first published by Shannon (1948).

Lotka (1926) was one of the first to examine the properties of power laws in scholarly communication. Interestingly, it took 13 years before the first citation to this work appeared, and this was a citation by Lotka himself! Historically, the third reference to Lotka (1926) in this set was made by Zipf (1948) in a work that lays out another important "law" in information science: Zipf's law. This law is somewhat similar to Lotka's law, and was likely inspired by it.

Interest in power-law distributions increased in the 2000s with the research on so-called scale-free distributions (Barabási & Albert, 1999) and processes such as "preferential attachments" (Barabási *et al.*, 2002). This process has also been discussed by Price (1976) as a "general theory of bibliometric and other cumulative advantage processes," and it builds on Lotka's work (as do, for example, Simon [1955] and Ijiri & Simon [1977], in still other research communities). Perhaps, Lotka (1926) can legitimately be considered as the founder of *iMetrics* as an intellectual program.

Institutionalization followed much later after the codification of this program in the early 1960s. However, let us first turn to 1934. In this year, the British mathematician and documentalist Samuel Bradford published his Law of Scattering that—informed by Lotka's law—states that papers in a specific subject will be concentrated in a limited number of journals, and that returns diminish rapidly when searching for these papers beyond this small set. This paper also built on a paper in *Science* by Gross and Gross (1927) discussing the same problem. Although cited in all three venues, this earlier paper was not cited as much as Bradford's (1934) contribution.

Garfield (1971) generalized Bradford's Law of Scattering into his Law of Concentration stating that the core of one set is in the tail of another. Because of this entanglement, a focus in the Web-of-Science on a set of a few thousand journals would sufficiently cover the whole (relevant) scientific literature. In other words, Lotka's and Bradford's laws provide us with mathematical expectations about the structure of bibliometric data. These formal insights enable us to focus the research process in *iMetrics*, but they need to be supplemented with empirical insights which became available from the early 1960s onwards in the history and philosophy of science.

Before the 1960s, study of the intellectual organization of the sciences was dominated by the history and philosophy of science (see, e.g., Lakatos & Musgrave, 1970). Kuhn's noted book (1962) opened the domain to empirical studies of the knowledge contents such as those pursued in the sociology of scientific knowledge (Barnes & Dolby, 1970). Using RPYS, however, Kuhn's oeuvre itself is not visible among the major foundational works for the *iMetrics* field.

Another author whom one would expect to be visible in this context is Robert K. Merton, who published on the sociology of science since his thesis in 1938 (Merton, 1938 and 1942). He is present in the 1957 peak in the RPYS of *Scientometrics* with his paper on priorities in discovery in science published in the *American Sociological Review* (Merton, 1957a). This paper was cited 19 times among the 167 cited references to publications in this year (11.4%). Furthermore, his better known work, a paper on the Matthew effect in science, is present in the 1968 peak in the spectrogram of *JoI,* and is cited 45 times among the 1,223 references to publications in this year (3.7%). The focus on these papers shows that books are not (or at least less systematically) included among the cited references. For example, Merton's important monograph *Social Theory and Social Structure* (1957b) was cited only four times in this same set.

The second major foundational publication—apart from Lotka (1926), present in all three venues—is the book *Little Science, Big Science* (1963), a citation classic by the historian Derek J. de Solla Price. In this book, Price articulated the exponential growth of science using the metaphor that in each generation more than half of all scholars are currently alive. With 32.3% of the references, this book has become more a concept-symbol than Price's earlier book *Science since Babylon* (1961) which was cited 30 times (14.8%) among the 203 references to publications in 1961. Price contributed another foundational work, a paper in *Science* entitled "Networks of scientific papers" (Price, 1965). While this work has the highest prominence in the papers published in *JASIST*-I, it was cited 160 times (in all three venues) out of a total of 624 references to this year (25.6%). In our opinion, these various studies by Price have had a shaping role in the development of *iMetrics*, as was already suggested by other studies (e.g., Wouters & Leydesdorff, 1994).

The year 1963 is indicated as highly significant for the development of *iMetrics*. The second largest contribution to the references constituting the peak of 1963 is Kessler's study entitled *Bibliographic Coupling between Scientific Papers*, published in *American Documentation*. It is credited with 82 (12.4%) of the 659 references to papers in 1963. In addition to being a valuable concept and retrieval technique in itself, bibliographic coupling provided a model that was used for developing the later concept of co-citation (Marshakova, 1973; Small, 1973). These two techniques have been influential both for information retrieval and for science mapping efforts (e.g., White & McCain, 1998).

Two papers by Eugene Garfield, the inventor of the Science Citation Index and the founder of the Institute of Scientific Information (1955; nowadays incorporated into Thomson Reuters), complement the peak in 1963. Both papers were published in *American Documentation*. Garfield and Sher (1963) obtained 41 references; Garfield (1963) was cited 17 times. Both papers argue in favor of using bibliographic instruments for historiography and have therefore been foundational to the research tradition to which our paper belongs. In addition, Garfield's original paper in *Science* about citation indexing (Garfield, 1955) was also included in one of the top three peaks of *JoI*.

Kaplan (1965) provided significant foundational work on the norms of citation behavior. This perspective was very prominent in *JASIST-I*, but less so in the other two venues. This may have to do with the focus of *JASIST-I* on topics related to scholarly communication (Milojević & Leydesdorff, 2013). In science and technology studies, the constructivist tradition (Gilbert, 1977; Cozzens, 1989) provided an alternative theory of citation (Luukkonen, 1997; cf. Cronin, 1984; Leydesdorff, 1998). Another reference that comes unexpectedly to the fore is Farrell's (1957) study entitled "The measurement of productive efficiency" in the *Journal of the Royal Statistical Society. Series A (General),* showing with other references the origin of contributions to *Scientometrics* from the side of mathematical statisticians.

The results of the RPYSs show that the intellectual program of *iMetrics* originates from a few precursors in the 1920s to the 1950s—notably, Alfred J. Lotka's publication in 1926 and Samuel C. Bradford's in 1934. However, 1963 seems to mark the year when the different research lines in the history of science (Derek de Solla Price), documentation (Michael M. Kessler), and citation indexing (Eugene Garfield) merged into this single enterprise. After 1963, the number of cited references changed by an order of magnitude. The peak in 1963 indicates the beginning of *iMetrics* as an increasingly coherent research program that would become institutionalized during the 1970s and 1980s (Lucio-Arias & Leydesdorff, 2009; Milojević & Leydesdorff, 2013; Van den Besselaar, 2001). *iMetrics* branches off from qualitative science and technology studies during the 1990s (Leydesdorff & Van den Besselaar, 1997) and joins the research programs in the information and library sciences increasingly during the 1990s and 2000s. A further stimulus to the growth of this specialty was provided by the focus on ranking introduced by Chinese participation (e.g., ARWU, 2004).

## 5    Discussion and conclusion

Let us now return to the question of the value of this historiography in relation to writing the intellectual history of a field. In our opinion, one should distinguish carefully between a reconstruction based on the perception by current authors (as in this study) and an intellectual history based on reading source materials. As White & McCain (1998, at p. 321) already formulated: "All ACA—that is, author co-citation analysis—can do, for the historian of ideas or any other party, is to identify influential authors and display their interrelationships from the citation record. It is no substitute for extensive reading and fine-grained content analysis, if someone is truly interested in the intellectual history of a field."

By focusing on citations, one risks to develop a Whiggish perspective on the history of science. Intellectual influences are not always manifest in references. For example, we noted that Kuhn's (1962) study, which was seminal to the development of science studies, could not be shown as influential in our set(s) using RPYS. Among practitioners certain references may be codified more than others, and important foundational work can be "obliterated by incorporation" (Garfield, 1975). For example, one no longer has to cite "Kuhn (1962)" when arguing about paradigms or paradigmatic developments. "Shannon and Weaver (1949)" is used more as a reference in *JASIST* than "Shannon (1948)," whereas the earlier publication contained the original knowledge claim.

In other words, we are reluctant to share White and McCain's (1998: 351) conclusion that author co-citation (ACA) "can be used to validate claims by historians and commentators." Like ACA, RPYS measures the (current) usage of references in scholarly manuscripts that primarily contain knowledge claims that have to be made convincingly. In addition to intellectual influence, referencing also has a function in the persuasion (Gilbert, 1977), and beyond this, referencing may also be strategic (Cozzens, 1989; Luukkonen, 1997; cf. Amsterdamska & Leydesdorff, 1989; Bornmann & Daniel, 2008; Leydesdorff & Amsterdamska, 1990). Authors may not even be aware of strategic citation behavior when focusing on the current state of the discussion—without sufficient knowledge of its history—or when citing themselves as (re-)inventors of historical breakthroughs (Althusser, 1974).

The above is not meant to devalue RPYS, but to define its use in reconstructing the perceived history of a field of science. As the American philosopher Alfred North Whitehead (1916) formulated: "A science which hesitates to forget its founders is lost." Using RPYS and other means of algorithmic historiography, one can map the aggregated reference behavior in a set and therewith the appreciation of its history by the carrying community or communities. Thus, the historiography of the field is reconstructed reflexively (Fujigaki, 1997). This reconstruction in practices can be considered as functional to the advancement of the sciences and the self-organizing dynamics of the scholarly discourses. The citation impact of a given paper is thus relative to the evolution of and the evaluation in the field(s) in which it is referenced.

We envisage that in the future more and richer means will be made available for the algorithmic reconstruction of the sciences. Even when these means are themselves dynamic, such as those based on moving averages (e.g., Leydesdorff, 2010; Leydesdorff & Goldstone, in press), in our opinion, one should keep in mind the difference between the history of the representation and the history of what is represented. The tension between these two histories provides a domain for methodological reflection on the epistemological status of the various representations.